\documentclass{moriond}

\def\be{\begin{equation}}
\def\ee{\end{equation}}
\def\bea{\begin{eqnarray}}
\def\eea{\end{eqnarray}}

\newcommand{\Photo}{}

\begin{document}
\vspace*{4cm}
\title{Testing regularity of black holes with X-rays and Gravitational Waves}

\author{ Swarnim Shashank \footnote{On behalf of: S. Riaz, R. Roy, A. B. Abdikamalov, D. Ayzenberg, C. Bambi, Z. Zhang, M. Zhou}}

\address{Center for Field Theory and Particle Physics and Department of Physics, Fudan University,\\ 200438 Shanghai, China}

\maketitle\abstracts{
Physically relevant solutions in general relativity often contain spacetime singularities, which are typically interpreted as a sign of breakdown of the theory at high densities/curvatures. Hence, there has been a growing interest in exploring phenomenological scenarios that describe singularity-free black holes, gravitational collapses, and cosmological models. We examine the metric put forth by Mazza, Franzin \& Liberati for a rotating regular black hole and estimate the regularization parameter $l$ based on existing X-ray and gravitational wave data for black holes. When $l=0$, the solution corresponds to the singular Kerr solution of general relativity, while a non-zero value of $l$ yields a regular black hole or a regular wormhole. The analysis reveals that the available data support a value of $l$ that is close to zero.}

\section{The BH mimicker metric}\label{sec:metric}

The metric is stationary, axisymmetric, and asymptotically flat, and its line element in the Boyer-Lindquist coordinates is written as~\cite{unka}
\begin{equation}
\label{line_element}
 {\rm d}s^2 = -\left(1 - \frac{2 M \sqrt{r^2 + l^2}}{\Sigma}\right){\rm d}t^2 + \frac{\Sigma}{\Delta}{\rm d}r^2 + \Sigma {\rm d}\theta^2 \\
 - \frac{4 M a \sin^2\theta \sqrt{r^2 + l^2}}{\Sigma}{\rm d}t{\rm d}\phi + \frac{A \sin^2\theta}{\Sigma}{\rm d}\phi^2 \,
\end{equation}
where,
\begin{equation}
\Sigma = r^2 + l^2+ a^2 \cos^2\theta \, , \quad
\Delta = r^2 + l^2 + a^2  - 2 M \sqrt{r^2 + l^2}  \, , \quad
A = \left(r^2+a^2 + l^2 \right)-\Delta a^2 \sin^2\theta \, .  
\end{equation}

The Horizon in the metric is expressed as,
\begin{equation}
r_{\rm H \pm} = \sqrt{ \left( M \pm \sqrt{M^2 - a^2} \right)^2 - l^2 }, 
\end{equation}

\section{Constraints from X-rays}\label{sec:xray}

For the case of X-ray data, we analyzed the 2019 \textsl{NuSTAR} observation of the Galactic black hole EXO~1846--031. The data from \textsl{NuSTAR} observation is quite good and suitable for such analysis. The constraint on the regularization parameter we find is,
\be\label{eq-f-c-x}
l/M < 0.49 \quad \textnormal{(90\% CL, only statistical error)} .
\ee
Full details can be found in Ref.~\cite{apna}.

\section{Gravitational Wave Constraints}\label{sec:gw}

Considering the equatorial geodesics of massive particles around a black hole, one can find the gravitational wave phase in leading order which for the regularization parameter $l^2$ is 2PN.
    \begin{equation}\label{eq:GW_phase_delta1}
        \Psi_{\mathrm{GW}}(f)=\Psi_{\mathrm{GW}}^{\mathrm{GR}}(f)-\frac{75}{32} u^{-1/3} \eta^{-4/5} l^2 + \mathcal{O}[l^4, u^0] \, .
    \end{equation}%
From the parameterized post-Einsteinian (ppE) formalism one can find the relation,
    \begin{equation}\label{eq:d1-p4}
        l^2 = -\frac{1}{100} \varphi_{4} \delta \varphi_{4} \, ,
    \end{equation}
where $\varphi_4$ is the PN phase for 2PN and $\delta \varphi_4$ is the correction to the phase used by the LIGO-Virgo Collaboration. Full formalism is provided in Ref.~\cite{apna}. Tab.~\ref{tab:gw} lists the constraints obtained for various GW events.

\begin{table}
    \centering
    \def\arraystretch{1.5}
    \setlength{\tabcolsep}{10pt}
    \begin{tabular}{|ccc|}
    \hline
        Event & $l/M$ (IMRPhenomPv2) & $l/M$ (SEOBNRv4P) \\
        \hline
		GW151226 & $0.50 ^{+ 0.26 }$ & $0.51 ^{+ 0.31 }$ \\
		GW170608 & $0.49 ^{+ 0.26 }$ & $0.57 ^{+ 0.29 }$ \\		
		GW190408A & $0.58 ^{+ 0.39 }$ & $0.66 ^{+ 0.53 }$ \\
		GW190412A & $0.57 ^{+ 0.27 }$ & $0.59 ^{+ 0.26 }$ \\
		GW190630A & $0.48 ^{+ 0.38 }$ & $0.47 ^{+ 0.31 }$ \\
		GW190707A & $0.44 ^{+ 0.24 }$ & $0.44 ^{+ 0.28 }$ \\
		GW190708A & $0.47 ^{+ 0.28 }$ & $0.45 ^{+ 0.28 }$ \\
		GW190720A & $0.45 ^{+ 0.29 }$ & $0.48 ^{+ 0.38 }$ \\
		GW190728A & $0.41 ^{+ 0.36 }$ & $--$ \\
		GW190828B & $0.46 ^{+ 0.33 }$ & $0.57 ^{+ 0.52 }$ \\
		\hline
    \end{tabular}
    \caption{Median and the $90$th percentile of the parameter $l/M$ from the significant binary black hole events in GWTC-1 and GWTC-2. 
    We only consider events with a total redshifted mass below 40~$M_\odot$ as we use only the inspiral phase for our analysis.
    Note the lower limit for the data is set to zero.}
    \label{tab:gw}
\end{table}

\section*{Acknowledgments}

I would like to thank C. Bambi and S. Riaz for their comments. I acknowledge the support from the China Scholarship Council (CSC), Grant No.~2020GXZ016646. The trip for Moriond was partially supported by the China Postdoctoral Science Foundation, Grant No. 2022M720035.

\end{document}